\documentclass[structabstract]{aa}

\usepackage{graphicx}
\usepackage{txfonts}
\usepackage{natbib}
\usepackage{subfigure}

\usepackage{latexsym,amssymb}
\usepackage{amsmath}
\usepackage{ucs}

\newcommand{\Msun}{M$_\odot$}

\begin{document}

   \title{CALIFA reveals prolate rotation in massive early-type galaxies:\\
A polar galaxy merger origin?}

\titlerunning{Prolate rotators in CALIFA - a polar merger origin?}
   
   \author{A. Tsatsi\inst{1}
   \and
   M. Lyubenova\inst{2}
   \and
   G. van de Ven\inst{1}
   \and
   J. Chang\inst{3}
   \and
   J. A. L. Aguerri\inst{4,5}
   \and
   J. Falc\'on-Barroso\inst{4,5}
   \and
   A.V. Macci\`o\inst{6,1}
   }
   
   \institute{Max-Planck-Institut f\"ur Astronomie, K\"onigstuhl 17, 69117 Heidelberg, Germany \email{tsatsi@mpia.de}
               \and
     Kapteyn Astronomical Institute, University of Groningen, Postbus 800, NL-9700 AV Groningen, the Netherlands
	\and
     Purple Mountain Observatory, the Partner Group of MPI f\"ur Astronomie, 2 West Beijing Road, Nanjing 210008, China
      \and
      Instituto de Astrof\'isica de Canarias, V\'ia L\'actea s/n, E-38205 La Laguna, Tenerife, Spain
      \and
      Departamento de Astrof\'isica, Universidad de La Laguna, E-38205 La Laguna, Tenerife, Spain
      \and
      New York University Abu Dhabi, PO Box 129188, Saadiyat Island, Abu Dhabi, United Arab Emirates
      }

    \date{}

  \abstract {We present new evidence for eight early-type galaxies (ETGs) from the CALIFA Survey that show clear rotation around their major photometric axis (``prolate rotation''). These are \object{LSBCF560-04}, \object{NGC 0647},  \object{NGC 0810}, \object{NGC 2484}, \object{NGC 4874}, \object{NGC 5216}, \object{NGC 6173} and \object{NGC 6338}. Including \object{NGC 5485}, a known case of an ETG with stellar prolate rotation, as well as \object{UGC 10695}, a further possible candidate for prolate rotation, we report ten CALIFA galaxies in total that show evidence for such a feature in their stellar kinematics. Prolate rotators correspond to $\sim$9$\%$ of the volume-corrected sample of CALIFA ETGs, a fraction much higher than previously reported. We find that prolate rotation is more common among the most massive ETGs. We investigate the implications of these findings by studying $N$-body merger simulations, and show that a prolate ETG with rotation around its major axis could be the result of a major polar merger, with the amplitude of prolate rotation depending on the initial bulge-to-total stellar mass ratio of its progenitor galaxies. Additionally, we find that prolate ETGs resulting from this formation scenario show a correlation between their stellar line-of-sight velocity and higher order moment $h_3$, opposite to typical oblate ETGs, as well as a double peak of their stellar velocity dispersion along their minor axis. Finally, we investigate the origin of prolate rotation in polar galaxy merger remnants. Our findings suggest that prolate rotation in massive ETGs might be more common than previously expected, and can help towards a better understanding of their dynamical structure and formation origin.}


\maketitle


\section{Introduction}
\label{sec:intro}

More than 60 years have passed since \cite{Contopoulos_1956} suggested that the intrinsic shape of elliptical galaxies could be triaxial. Since then it has been established that the existence and persistence of triaxial galaxies is theoretically permitted \citep[e.g.][]{Aarseth_1978, Binney_1985}, and that if elliptical, or in general, early-type galaxies (ETGs) are such systems, then they are expected to show two types of stable stellar rotation; rotation around their short axis, as in the typical case of oblate systems (oblate rotation), as well as rotation around their long axis, so called ``prolate rotation"\footnote{Note: ``Prolate rotation'' is often referred to as ``long axis" rotation, or ``minor axis'' rotation, the latter meaning a velocity gradient along their projected minor axis.}. 

 This suggests that any observations of an undisturbed early-type galaxy that shows rotation around its major apparent axis is indicative of its triaxial shape. However, several attempts to find such systems have been unsuccessful in the past \citep[e.g.][]{Bertola_1988} and until now, only a few observations of galaxies with clear prolate rotation exist. These cases concern: NGC 1052 \citep{Schechter_1979, Davies_Illingworth_1986}, NGC 4406, NGC 5982, NGC 7052, NGC 4365, NGC 5485 \citep{Wagner_1988}, NGC 4261 \citep{Davies_Birkinshaw_1986, Wagner_1988}, NGC 4589 \citep{Wagner_1988, Moellenhoff_1989}, AM 0609-331 \citep{Moellenhoff_1986} and \object{M87} \citep{Davies_Birkinshaw_1988, Emsellem_2014}, as well as NGC 5557 in \citealt{Krajnovic_2011}, totalling 11 objects\footnote{Note: There is also evidence for (although not clear) prolate rotation in NGC 2749, IC 179 \citep{Jedrzejewski_1989} NGC 3923 \citep{Carter_1998} and NGC 7626 \citep{Davies_Birkinshaw_1988}.}. Most of these prolate rotators belong in the potential wells of galaxy groups or clusters, while 3 of them, NGC 4589, NGC 5485 and AM 0609-331 show strong dust lanes along their minor axes, providing an additional visual evidence of triaxiality.
Only for 6 of the above 11 known prolate rotating ETGs, two-dimensional Integral Field Unit (IFU) spectroscopy of the stellar kinematics has been carried out so far: NGC 4261, NGC 5485, NGC 4365, NGC 4406, NGC 5557 and \object{M87} \citep{Davies_2001, Emsellem_2004, Krajnovic_2011, Emsellem_2014}. Their studies have revealed that prolate rotation may often coexist with oblate rotation, often in the form of a kinematically decoupled component (KDC), a central stellar component with distinct kinematic properties from those of the main body of the galaxy, often found to reside in many ETGs \citep[e.g.][]{Mcdermid_2006, Krajnovic_2011}.

While the dynamical stability of prolate rotation in triaxial galaxies has been extensively studied theoretically, constraining the formation origin of such systems is yet a highly challenging task, mainly due to the fact that only few observations of prolate-rotating ETGs exist in the literature.

A growing amount of evidence from cosmological simulations suggests that massive ETGs of stellar mass $M_{*}\gtrsim$ \ensuremath{10^{11}$\Msun$ } have been assembled in two phases: (i) an early rapid dissipational formation phase (such as a gas-rich major merger), taking place at redshift z\textgreater2 and contributing to the main build-up of the central 1-2 effective radii ($r_e$) of the galaxy, followed by a (ii) second phase of satellite accretion (gas-poor minor mergers) that built up their outer parts (r \textgreater$1-2r_e$) at z\textless2 \citep[e.g.][]{Vitvitska_2002, Khochfar_2009, Johansson_2012, Lackner_2012, Naab_2014}. This ``two-phase'' assembly formation scenario is in line with a series of observational lookback studies that report a significant growth in mass and in size of massive ETGs at radii $r$\textgreater$1-2r_e$ since z$\sim$2 \citep[e.g.][]{Zirm_2007, vanDokkum_2008, vanDokkum_2010, Perez_2013, Burke_2013, vdWel_2014}. It is also supported by observations of ETGs luminosity function evolution \citep[e.g.][]{Bell_2004, Faber_2007}, as well as wide-field observations of the stellar kinematics of many ETGs that show central, oblate rotating components within $1-2r_e$, consistent with being formed through an initial gas rich major merger, embedded in more spherical and slowly rotating structures dominating at larger radii ($2-4r_e$), consistent with a late satellite accretion assembly phase \citep{Arnold_2014}. 

However, in the case of ETGs that show strong prolate rotation in their central regions ($r$\textless$1-2r_e$), it is not clear whether and how the above picture of an early and rapid dissipational process could form a triaxial inner structure with no (significant) oblate rotation.
It is possible that the two-phase assembly scenario, where the central region was formed by a major merger still holds for this special case of prolate rotators. This is supported by the study of \cite{Naab_2003, Jesseit_2007}, who showed that the prolate rotation in the remnant of a 1:1 merger is stronger than in unequal-mass merger remnants. \cite{Cox_2006} however, showed that dissipational, equal-mass disk mergers result in merger remnants with stronger oblate rotation. \cite{Bois_2011}, comparing the kinematics of merger remnants with the observed kinematics of ETGs, also showed that slow rotators formed in 1:1 mergers can show significant kinematic misalignments, with strong prolate rotation in the case of gas-poor merger remnants \citep{Hoffman_2010}.

Given the motivations above, it is most likely that the inner regions of present-day massive ETGs that show strong prolate rotation and no (significant) oblate rotation, may have been formed preferentially by gas-poor (dry) major mergers. However, it is yet not clear the reason why only few observations of such systems exist so far in the literature.

\begin{table*}
\begin{center}
\begin{tabular}{llcccc}
\hline\hline
Name & T/G & d & M$_*$ & r$_e$ & $\Psi$ \\
 &  & (Mpc) &  (\ensuremath{10^{11}$\Msun$}) & (kpc)  & (deg) \\
\hline
LSBCF560-04 &E5/BCG & 238 & $7.1\pm0.6$\tablefootmark{(a)} &   17.9  & 57$\pm$5  \\
NGC 0647&E7/in a group & 184 & $3.7 \pm 0.3$\tablefootmark{(a)} & 8.7   & 72$\pm$3 \\
NGC 0810 &E5/in a pair &110 & 2.12$\substack{+0.16 \\ -0.08}$\tablefootmark{(b)} & 9.3    &87$\pm$2   \\
NGC 2484&E4/BCG & 192 & $5.0 \pm 0.5$\tablefootmark{(a)} &  12.5  &52$\pm$4   \\
NGC 4874&E0/BCG & 102 & 2.8$\substack{+1.6 \\ -1.1}$\tablefootmark{(b)} & 12.4   & 86$\pm$5 \\
NGC 5216 &E0/in a pair &42 & 0.230$\substack{+0.02 \\ -0.001}$\tablefootmark{(b)} &4.1 & 66$\pm$5   \\
NGC 5485 &E5/in a group &27 & 0.9$\substack{+0.1 \\ -0.4}$\tablefootmark{(b)} &4.1  &  80$\pm$3  \\
NGC 6173&E6/BCG & 126 & 2.5$\substack{+0.3 \\ -0.2}$\tablefootmark{(b)} &  30.5   & 80$\pm$2    \\
NGC 6338&E5/BCG & 117 & 3.0$\substack{+0.3 \\ -0.7}$\tablefootmark{(b)} &  17.0   & 36$\pm$4  \\
UGC 10695$\tablefootmark{*}$&E5/in a group  & 120  &  2.7$\pm$0.2\tablefootmark{(a)} &  15.7  & 87$\pm$2 \\
\hline
\end{tabular}
\end{center}
\begin{center}
\caption{\small{Properties of the CALIFA prolate rotators: Col.1: Name. Col.2: Hubble type T, as in \cite{Walcher_2014}, and group membership G (NED/SIMBAD). Col.3: Redshift-based distance in Mpc. Col.4: Available stellar mass estimates,\tablefoottext{a}{from WISE photometry by \cite{Norris_2016},} and \tablefoottext{b}{from CALIFA DR3, as in \cite{Walcher_2014}.} Col.5: Effective radius, determined by growth curve analysis of SDSS images of each galaxy, as in \cite{Walcher_2014}. Col.6: Global kinematic misalignment angle $\Psi$. Cosmological angular size distances are calculated adopting H$_0$ = 70 km$\cdot$s$^{-1}$$\cdot$Mpc$^{-1}$, $\Omega_m$=0.3, $\Omega_{\Lambda}$=0.7.
\tablefoottext{*}{\object{UGC 10695} is listed as a further possible candidate, showing evidence for prolate rotation.}} }
\label{tab:califatab}
\end{center}
\end{table*}

Motivated by all the above evidence, we have made use of the Calar Alto Legacy Integral Field Area (CALIFA) survey \citep{Sanchez_2012}, which provides IFU data for a statistically well defined sample of $\sim$600 galaxies across the Hubble sequence, in order to search for possible prolate rotating ETGs. Here, we present 10, massive ($M_{*}\sim$ \ensuremath{10^{11}$\Msun$ }) ETGs that show prolate rotation in their stellar kinematics, including one known (NGC 5485), the discovery of eight new clear cases of massive prolate-rotating ETGs, as well as one further possible candidate (\object{UGC 10695}), adding a significant fraction to the cases of prolate rotation that exist so far in the literature.

Additionally, we investigate a possible merger origin of these systems by studying the kinematics of simulated ETGs formed in $N$-body simulations of major mergers in an observational-like fashion. We show that polar major mergers of disk galaxies can produce prolate-shaped merger remnants with strong rotation around their major axis, which could be the progenitors of the observed present-day massive prolate rotating ETGs. Such a formation scenario is in line with the existing picture of a two-phase assembly of massive ETGs and can explain the rarity of observations of prolate rotators as a natural consequence of the infrequency of dry major polar mergers.

This paper is organised as follows: in Section~\ref{sec:2} we describe the sample and present the observations of prolate rotators in the CALIFA Survey, in Section~\ref{sec:3} we describe the set of major merger simulations, the resulting shape and kinematics of the simulated remnant ETGs, and the origin of their prolate rotation. Finally, we conclude in Section~\ref{sec:4}.\\

\begin{figure*}
\center
\includegraphics[trim = 0.7cm 3.3cm 21.7cm 7.5cm,clip, width=16.5cm, ]{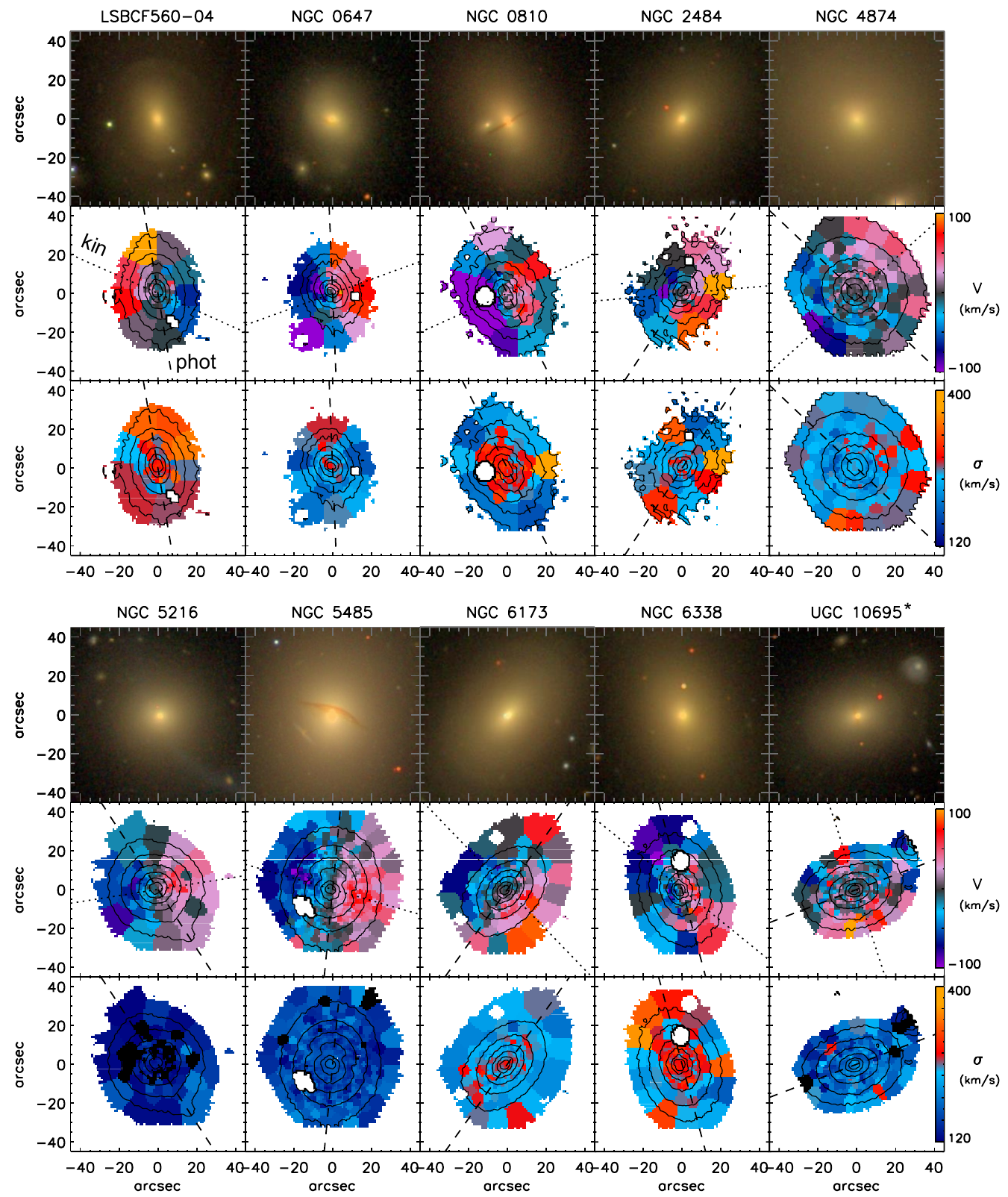}
\center
\caption{The CALIFA sample of prolate rotators. Top panel: \object{LSBCF560-04}, \object{NGC 0647},  \object{NGC 0810}, \object{NGC 2484} and  \object{NGC 4874}. The top row shows a color-composite SDSS image of each galaxy, while the second and third row show the stellar line of sight velocity V and velocity dispersion $\sigma$ in km/s, respectively, extracted from the V500 CALIFA dataset. The dashed line shows the photometric major axis, measured in the outer parts of each galaxy, using SDSS images. The dotted line shows the global kinematic axis, measured as in \citet[see Appendix C]{Krajnovic_2006}. Bottom panel: same as top panel, for \object{NGC 5216}, \object{NGC 5485}, \object{NGC 6173}, \object{NGC 6338}, and \object{UGC 10695}.}
\label{fig:figure1}
\end{figure*}

\section{Observations}
\label{sec:2}

The Calar Alto Legacy Integral Field Area (CALIFA) survey \citep{Sanchez_2012} provides IFU data for a statistically well defined sample of $\sim$600 galaxies across the Hubble sequence. The CALIFA sample is selected from a Mother Sample of 938 galaxies in the photometric catalogue of the 7th data release of the Sloan Digital Sky Survey (SDSS) \citep{Abazajian_2009}, with the main selection criteria being an angular isophotal diameter of 45$\arcsec$$\le$$D_{25}$$\le$80$\arcsec$, and a redshift range of 0.005 $\le$z$\le$0.03.

The survey uses the PPAK IFU \citep{Verheijen_2004, Kelz_2006} of the Potsdam Multi-Aperture Spectrograph, PMAS \citep{Roth_2005} at the 3.5 m telescope of CAHA, with a Field-of-View (FoV) that can extend up to several effective radii ($r_{e}$), using two different setups (V500 and V1200) of resolutions R$\sim$850 and R$\sim$1650, respectively. For a more detailed description of the CALIFA survey and extraction of stellar kinematics, see \cite{Walcher_2014, FalconBArroso_2016}.

In this work we focus on observations of 10 early-type galaxies (ETGs), chosen by visual inspection of the stellar kinematics of the CALIFA 3rd Data Release\footnote{See \cite{Sanchez_2016} for a detailed description of the CALIFA DR3.} (DR3) sample of 667 galaxies. These are: \object{LSBCF560-04}, \object{NGC 0647},  \object{NGC 0810}, \object{NGC 2484},  \object{NGC 4874}, \object{NGC 5216}, \object{NGC 5485}, \object{NGC 6173}, \object{NGC 6338} and \object{UGC 10695}. Out of these 10 galaxies, 7 belong to the CALIFA kinematic sub-sample of 81 ETGs (E+S0s), described in \cite{FalconBArroso_2016}, a statistically well-defined sample of $\sim$300 galaxies in total, while 3 (\object{LSBCF560-04}, \object{NGC 0647} and \object{NGC 2484}) belong to the CALIFA Extension Sample, an inhomogeneous sample of galaxies observed in the CALIFA set-up \citep{Sanchez_2016}. Some of their qualitative properties, including their Hubble types, group memberships, distances, stellar masses and effective radii are shown in Table~\ref{tab:califatab}.

We derived our stellar kinematics maps following the strategy described in \cite{FalconBArroso_2016}. However, there are a few differences that we adopted in our work. We used the CALIFA V500 data set only, as these observations typically reach fainter magnitudes and this way we are able to better recover the stellar kinematics in the outer parts of the galaxies. In order to reliably measure the line-of-sight velocity, velocity dispersion and the higher order Gauss-Hermite terms $h_3$ and $h_4$, we binned the data to S/N$\sim$40 using the Voronoi binning technique as implemented by \cite{Cappellari_2003}. Then we used a subset of $\sim$330 stars from the IndoUS library \citep[][the same set as in \citealt{FalconBArroso_2016}]{Valdes_2004}, and fitted the binned spectra in the wavelength range 4250.0 -- 5500.0 $\textrm{\AA}$. 

In this way we extracted the stellar kinematics shown in Figure~\ref{fig:figure1}. The CALIFA maps reveal prolate rotation for all 10 galaxies, of amplitudes ranging from $\sim$40 to 100 km/s. In all cases we find strong kinematic evidence for triaxiality, as the stellar rotation of the main body of the galaxy is prolate (around the major apparent axis). In the case of \object{UGC 10695} there is evidence for prolate rotation in its stellar line-of-sight velocity map (however not as clear as the other cases), and thus is listed as a possible further candidate in our sample. In the case of \object{NGC 6338} the rotation is prolate only in its inner parts, while the outer parts show oblate rotation-- a new evidence of an elliptical galaxy with a kinematically decoupled, prolate-rotating component.

In order to quantify the kinematic misalignment for our sample of galaxies, we estimate the global kinematic misalignment angle $\Psi$ as defined in \citet{Franx_1991}:

\begin{equation}
\label{Eq:T0}
sin\Psi=\mid sin(PA_{phot} - PA_{kin}) \mid
\end{equation}

where PA$_{phot}$ is the position angle of the photometric major axis, measured in the outer parts of each galaxy using SDSS images, measured as in \citet{Walcher_2014}, while PA$_{kin}$ is the global kinematic position angle measured from the CALIFA velocity maps of each galaxy, as in \citet[see Appendix C]{Krajnovic_2006}. Angle $\Psi$ as defined above, shows the projected misalignment between photometry and kinematics of a stellar system. For most ETGs this quantity is rather low, showing alignment typical for oblate systems, and for example in the ATLAS$^{3D}$ Survey approximately 90$\%$ of ETGs show $\Psi$\textless15$^{\circ}$ \citep{Krajnovic_2011}.

Table~\ref{tab:califatab} shows the derived global misalignment angles for all the galaxies in our sample. For most galaxies we find $\Psi$\textgreater50$^{\circ}$. For \object{NGC 6338} we find $\Psi$=36$^{\circ}\pm$4, as there is prolate rotation in the inner parts, while the apparent oblate rotation of the galaxy in the outer parts influences the value of global $\Psi$. UGC 10695, listed here as a possible candidate, shows $\Psi$=87$^{\circ}\pm$2, a further indication for prolate rotation.

It is interesting to note that almost all of the galaxies in our sample belong in groups or clusters. Five out of ten are the brightest galaxies in their clusters (BCGs), with relatively high stellar masses ($M_{*}\gtrsim$ \ensuremath{2\cdot10^{11}$\Msun$ }).

One can also see that two galaxies in our sample show strong dust lanes along their minor photometric axis in their SDSS images (Figure~\ref{fig:figure1}): \object{NGC 5485}, a known case of an ETG with prolate rotation and a minor axis dust lane \citep{Wagner_1988, Emsellem_2011}, and the new case of \object{NGC 0810}. Assuming that dust and gas settle in the principal planes of a galaxy, the existence of a minor axis dust lane is the visual evidence for triaxiality \citep{Bertola_1978, Merritt_1983}. We expect that the 10 prolate rotators of our sample must be, to some extent, triaxial or prolate. 

We find that prolate rotation, and the triaxiality in massive ETGs implied from this stellar kinematic feature, may not be as rare as previously thought. In order to estimate the rate of occurrence of this kinematic feature in ETGs, we consider the subset of 6 prolate rotators which belong to the statistically well defined CALIFA Kinematic Sub-sample of 81 ETGs \citep{FalconBArroso_2016}. After applying volume corrections for each galaxy, following \cite{Walcher_2014}, we report a fraction of $\sim$9$\%$ of prolate-rotating ETGs in the volume of the CALIFA Kinematic Sub-Sample ETGs\footnote{UGC 10695 is listed as a possible candidate and as such, is not included in this fraction. We have also excluded the rest of the galaxies from our sample (\object{LSBCF560-04}, \object{NGC 0647} and \object{NGC 2484}), which belong to the inhomogeneous collection of the CALIFA Extension Sample, and for which volume corrections cannot be applied.}. Notably, this fraction becomes $\sim$27$\%$ in the volume of ETGs with stellar masses $M_{*}\gtrsim$ \ensuremath{2\cdot10^{11}$\Msun$ }.

These fractions seem to agree with the corresponding ones from the ATLAS$^{3D}$ Survey, where 6 galaxies (\object{NGC 4261}, \object{4365}, \object{4406}, \object{5485}, \object{5557} and \object{M87}) were identified to show prolate rotation in their stellar kinematics. This yields 6 out 51 ETGs with prolate rotation in the mass range of ETGs with $M_{*}\gtrsim$ \ensuremath{10^{11}$\Msun$ }, 
a fraction of $\sim$12$\%$, which becomes $\sim$23$\%$ (5 out of 22) in the mass range of $M_{*}\gtrsim$ \ensuremath{2\cdot10^{11}$\Msun$ } of ATLAS$^{3D}$ \citep[D.Krajnovi\'c, E. Emsellem, priv. comm.,][]{Krajnovic_2011, Emsellem_2014}.

What are the implications of these findings for the formation of such systems? Could prolate rotators be the end-products of major mergers? Considering the rarity of observations of such systems so far, their formation origin is still poorly understood. According to the two phase-assembly scenario, the central parts of massive ETGs have been preferentially assembled through gas-rich major mergers approximately $\sim$10 Gyr ago. However we do not observe clear signs of oblate rotation in most of the CALIFA prolate rotators presented here. This suggests that for this special type of objects their merger origin might have been rather gas-poor. In what follows, we explore the dynamical structure of remnants resulting from such a formation scenario.\\


\section{Polar galaxy mergers}
\label{sec:3}

We present a possible formation origin of prolate rotation in massive early-type galaxies, by perfoming $N$-body simulations of major mergers of spiral galaxies in polar orbits.
\\
\subsection{Simulations}

 The binary merger simulations we use were performed using the TreeSPH-code GADGET-3\footnote{See \citealt{Springel_2005} for a detailed description of the previous version of this code, GADGET-2.}, as used in \cite{Chang_2013}. In order to investigate the gas-poor merger scenario of prolate galaxies, our simulations do not include cold or hot gas (hence our simulations are $N$-body).

 The two progenitor disk galaxies are identical and they are composed of a stellar disk and a stellar bulge, which are embedded in a dark matter halo. The disk component follows an exponential profile, while the stellar bulge and the dark matter halo of each progenitor follow a \cite{Hernquist_1990} profile, the same as described in \cite{Springel_dmH2005}. The disk of one of the progenitor galaxies is aligned with respect to the orbital plane, while the other progenitor has its disk inclined by 90$^{\circ}$ with respect to its companion and the orbital plane. The initial radial and tangential velocity of each progenitor is $\sim$360 and 180 km/s, respectively, and their initial distance is 325 kpc.
 
\begin{figure}[h!]
\center
\includegraphics[trim = 0.7cm 3.5cm 24cm 7.5cm,clip, width=9cm, ]{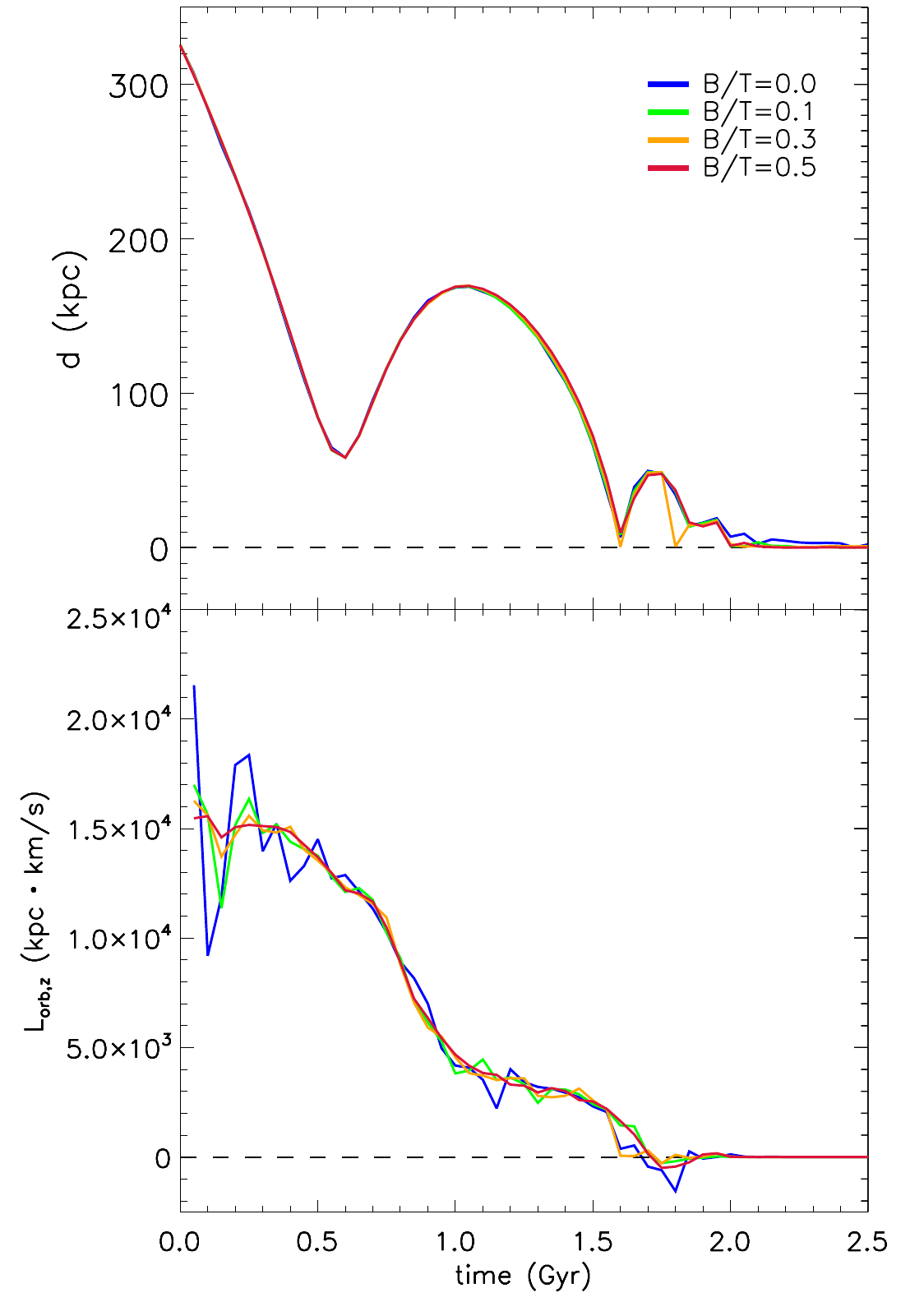}
\center
\caption{Relative distance d of the two progenitors (top) and orbital angular momentum L$_{orb,z}$ of one of the progenitors (bottom) as a function of time during each merger simulation. Different colors correspond to the different B/T ratios of each progenitor.}
\label{fig:figuretr}
\end{figure}

  We run 4 realizations of the simulation described above, with different bulge-to-total stellar mass ratios (B/T) of the progenitor galaxies. We adopt a range of B/T ratios of 0.0, 0.1, 0.3 and 0.5 (see Table~\ref{tab:Prolate_sim}). All the simulations have the same total number of stellar and dark matter particles (N=2\,425\,432), so that the each merger remnant has a stellar mass of \ensuremath{M_*=2.6 \times 10^{ 11 } $\Msun$ } and a dark matter halo of \ensuremath{M_{dm}=1.6 \times 10^{ 13 } $\Msun$ }. The softening length is 70 pc for stellar and 300 pc for dark matter particles, respectively. Each progenitor galaxy evolves initially in isolation for 2 Gyr, so that it exhibits a reasonably steady structure before the merger. After that the merger simulation starts and lasts for approximately 8 Gyr. Figure~\ref{fig:figuretr} shows the relative distance of the two progenitors, as well as their orbital angular momentum as a function of time during each merger simulation.\\
\begin{table}
\begin{center}
\begin{tabular}{rccccc}
\hline\hline
Name & Progenitor B/T & $r_{hm}$  & $b/a$ & $c/a$ & $T$  \\
&  & (kpc) & & &  \\
\hline
M0&0.00 & 16.00 & 0.57 &   0.51  &  0.91\\
M1&0.10 & 14.83 &0.57 &  0.47 & 0.87\\
M2&0.30 & 12.42 & 0.61 & 0.51 & 0.84\\
M3&0.50  & 9.90 & 0.63  &0.50 & 0.81\\
\hline
\end{tabular}
\end{center}
\begin{center}
\caption{Properties of the simulated remnants: Name, mass ratio between bulge and total mass of each progenitor B/T, half-mass radius $r_{hm}$, axial ratios $b/a$, $c/a$ and triaxiality parameter T (see Equation~\ref{Eq:T}) of the simulated remnants, estimated within $r_{hm}$.}
\label{tab:Prolate_sim}
\end{center}
\end{table}

\subsection{Extracting mock observational data}

In order to connect the intrinsic mass and orbital distribution of our simulated merger remnant galaxies with observable properties, we create two-dimensional mock stellar mass and stellar kinematic maps following the method described in \cite{Tsatsi_2015}.

 Stellar particles are projected along a chosen viewing angle and then binned on a regular grid centered on the baryonic center of mass of the galaxy. We adopt a grid size of 20 $\times$20 kpc$^2$ and a pixel size of 0.5 kpc, which corresponds to the spatial resolution of CALIFA ($\sim$1\arcsec), assuming that our simulated galaxies are observed at a distance of $\sim$100 Mpc. The size of the corresponding ``CALIFA-like'' field-of-view extends out to $\sim$1 or 2 half-mass radii of each remnant.
 
 The bulk velocity of the galaxy is estimated within a sphere of 50 kpc around the center and subtracted from all particle velocities. We then extract stellar mass-weighted line-of-sight kinematic maps for each of our merger remnants. The maps are spatially binned using the 2D Voronoi binning method \citep{Cappellari_2003}, based on a minimum number of particles per pixel in the map. Signal corresponds to the number of particles per pixel and we adopt Poisson noise, such that our signal-to-noise ratio per bin corresponds approximately to an average target value of $\ensuremath{S/N\sim35}$ for all the remnants. 
 
 The mass-weighted stellar line-of-sight velocity distribution (LOSVD) is then extracted for every Voronoi bin and fitted with the Gauss--Hermite series \citep{vdMarel_1993}, as implemented by \citet[see Appendix]{vdVen_2006} allowing us to retrieve the Gauss--Hermite parameters of the LOSVD (V, $\sigma$, $h_3$, $h_4$) of the final merger remnants.\\

  \subsection{Shape and kinematics of the simulated remnants}
The shape properties of the resulting merger remnants are shown in Table~\ref{tab:Prolate_sim}. The remnants become more compact as the B/T of their progenitors increases and their half-mass radius (r$_{hm}$) decreases.

\begin{figure}
\center
\includegraphics[trim = 3.5cm 11.5cm 16.5cm 6cm,clip, width=10cm, ]{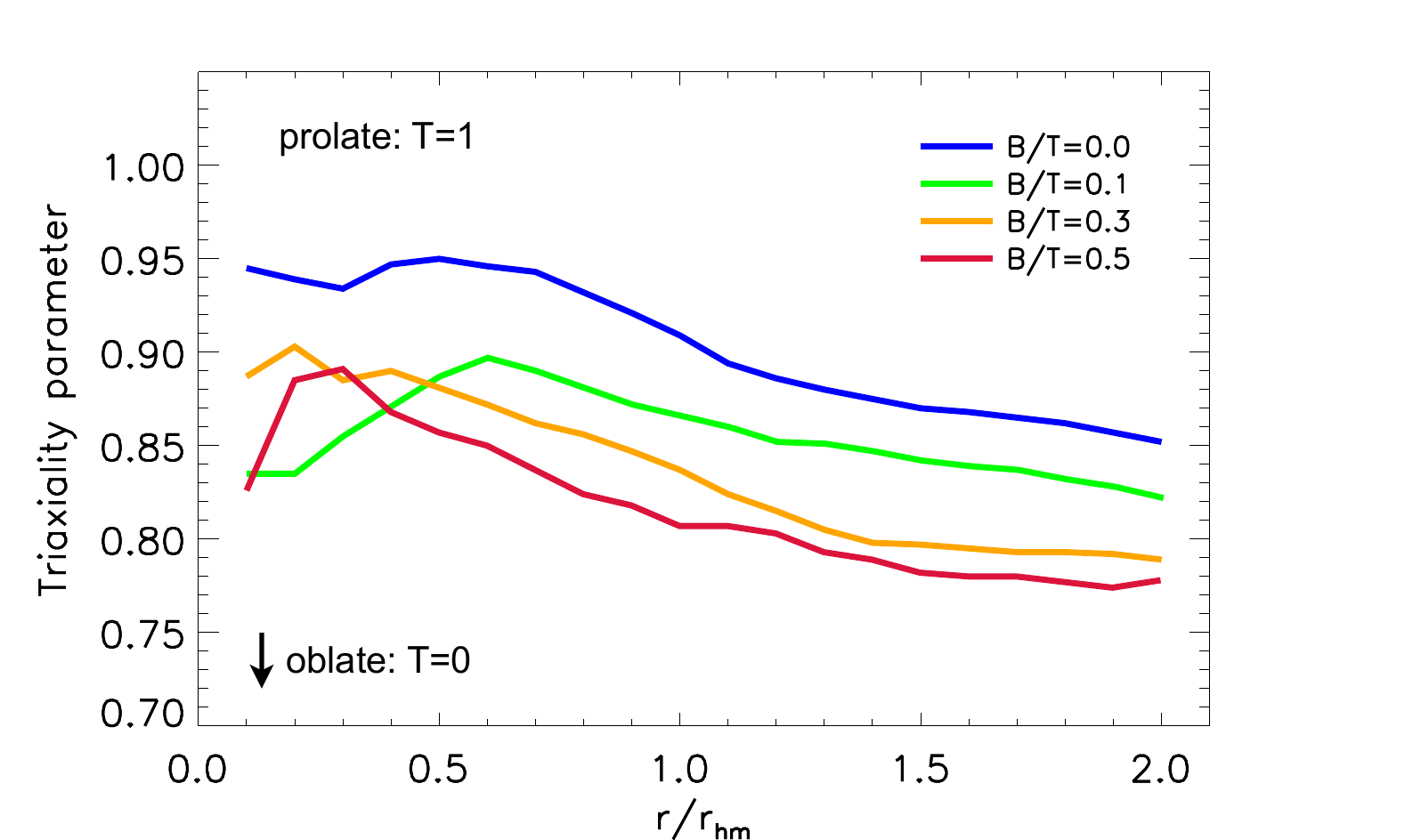}
\center
\caption{Triaxiality parameter T as a function of distance from the center of each remnant r/r$_{hm}$, for all the simulated merger remnants. Different colors correspond to the different B/T ratios of their progenitors.}
\label{fig:figure2}
\end{figure}

Figure~\ref{fig:figure2} shows, as a function of distance from the center of each remnant, the triaxiality parameter T, defined as:
\begin{equation}
\label{Eq:T}
T=\frac{1-(b/a)^2}{1-(c/a)^2}
\end{equation}
where a\textgreater b\textgreater c correspond to the principal axes of an ellipsoid, computed by extracting the eigenvalues of the moment of inertia tensor within spherical shells of radius r for each remnant (for oblate ellipsoids T=0 and for prolate ellipsoids T=1).

 We see that all our simulated remnants are highly prolate within two half-mass radii, with T$>$0.8. For radii r\textgreater 0.5r$_{hm}$ there appears to be a trend of lower T with increasing B/T ratio of the progenitors.
 
\begin{figure*}
\center
\includegraphics[trim = 0.5cm 15cm 21cm 2cm,clip, width=13cm, ]{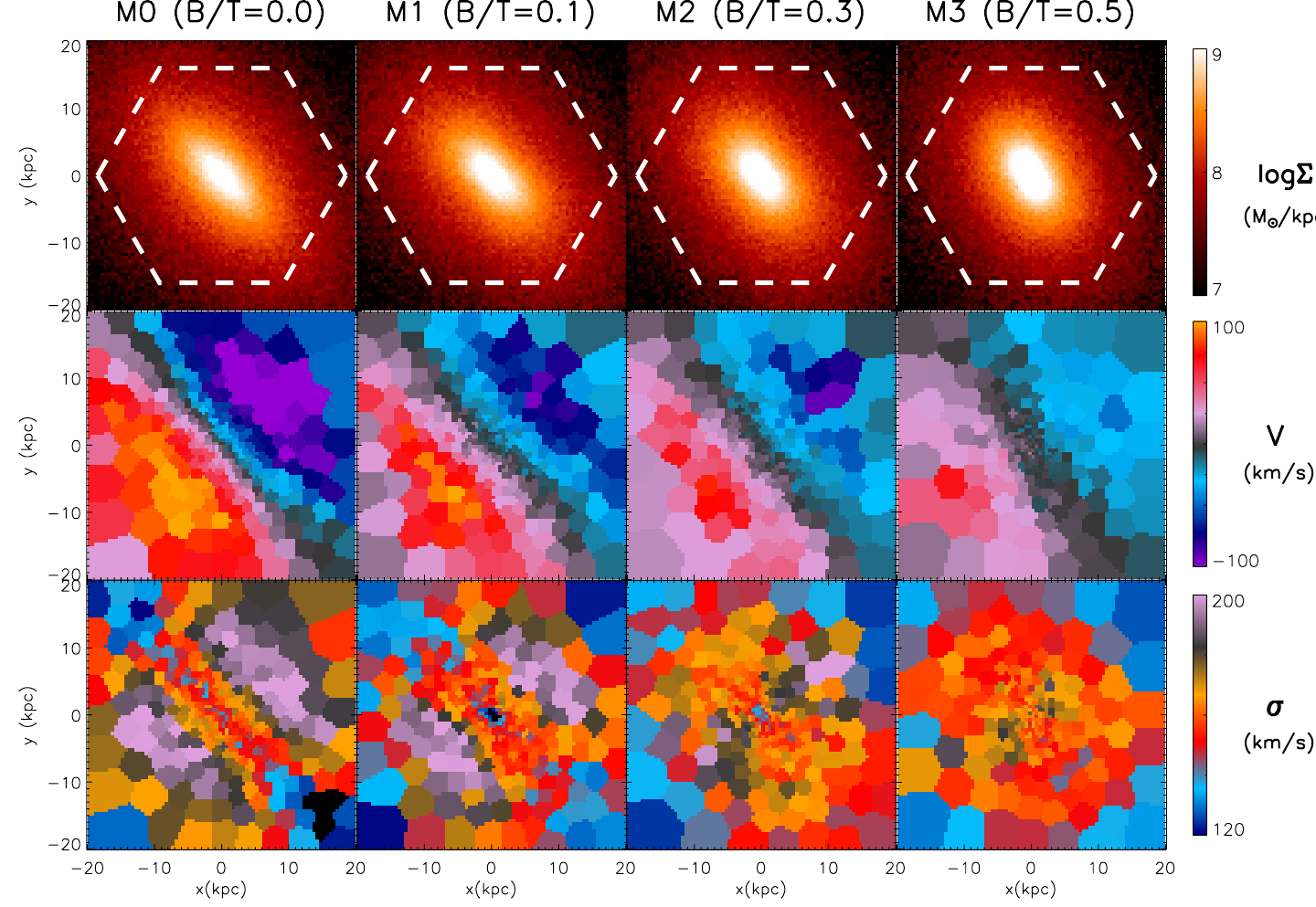}
\center
\caption{Stellar mass and line of sight orbital distribution of the simulated merger remnants M0-3 (the B/T ratio of their progenitors increases from left to right). The projected plane x-y corresponds to the orbital plane of the merger. From top to bottom: Mass distribution $log\Sigma$, V, and $\sigma$. The ``CALIFA-like'' hexagonal field of view for the assumed distance is overplotted on the mock images of the first row.}
\label{fig:figure3}
\end{figure*}

The projected stellar mass, mean velocity and velocity dispersion of all the final merger remnants are shown in Figure~\ref{fig:figure3}. All the  remnants show prolate rotation, with amplitude depending on the initial B/T ratio of their progenitors. A lower B/T ratio of the progenitors results in remnants with stronger prolate rotation, with a maximum amplitude of $\sim$100 km/s in the case of two bulgeless progenitors (B/T=0). 

\begin{figure*}
\center
\includegraphics[trim = 0.5cm 11.5cm 11cm 1cm,clip, width=16cm, ]{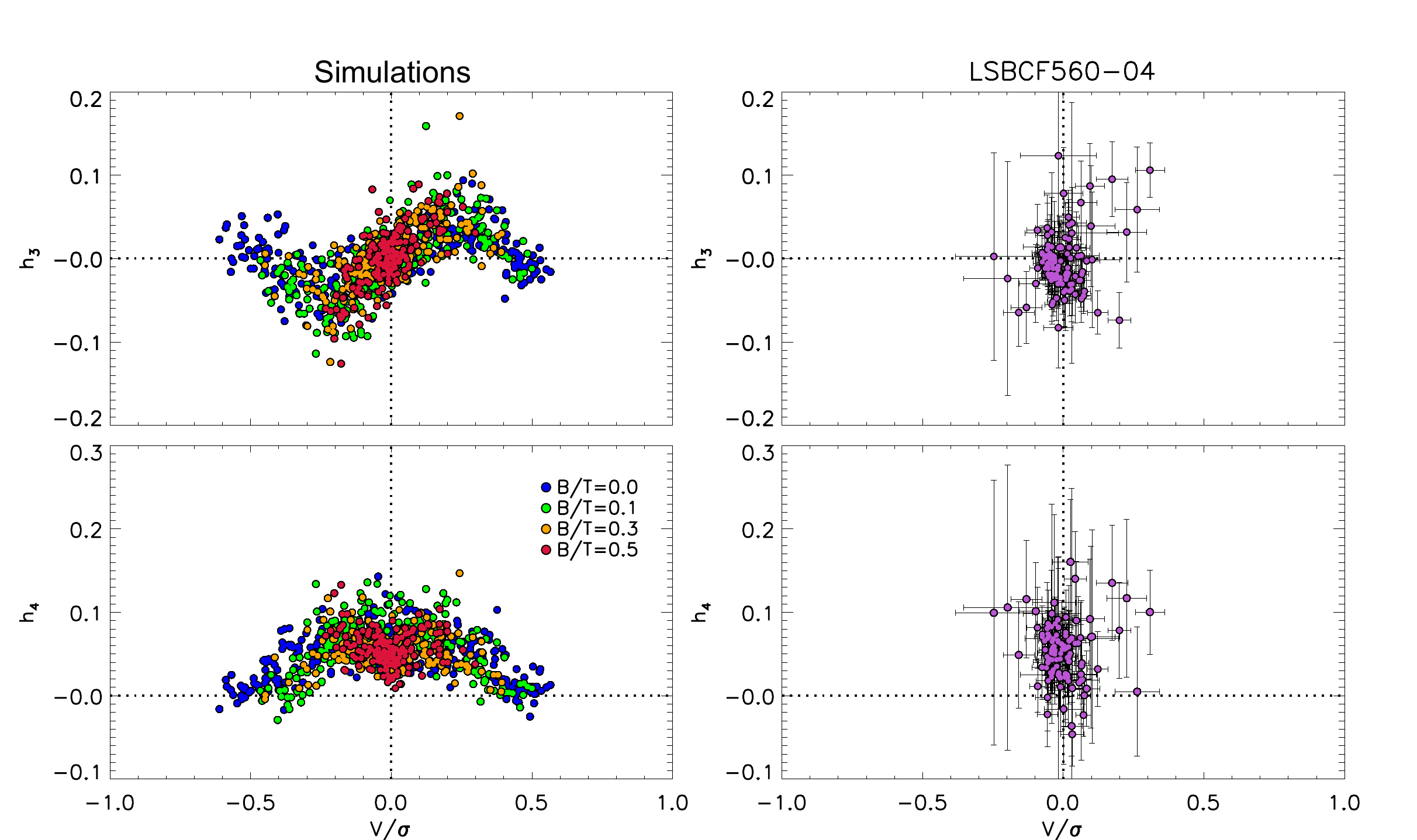}
\center
\caption{Left: Pixel values within $r_{hm}$ of $h_3$ (upper panel), $h_4$ (bottom panel) versus $V/\sigma$ extracted from the simulated kinematic maps for all the simulated remnants M0-M3. Different colors correspond to the different B/T ratios of their progenitors. Right: Same as left, for one of the CALIFA prolate-rotating galaxies, \object{LSBCF560-04}, that shows weak evidence for an $h_3$-$V/\sigma$ correlation.}
\label{fig:figure4}
\end{figure*}
\begin{figure*}
\center
\includegraphics[trim = 0.5cm 13cm 20.5cm 0cm,clip, width=13cm, ]{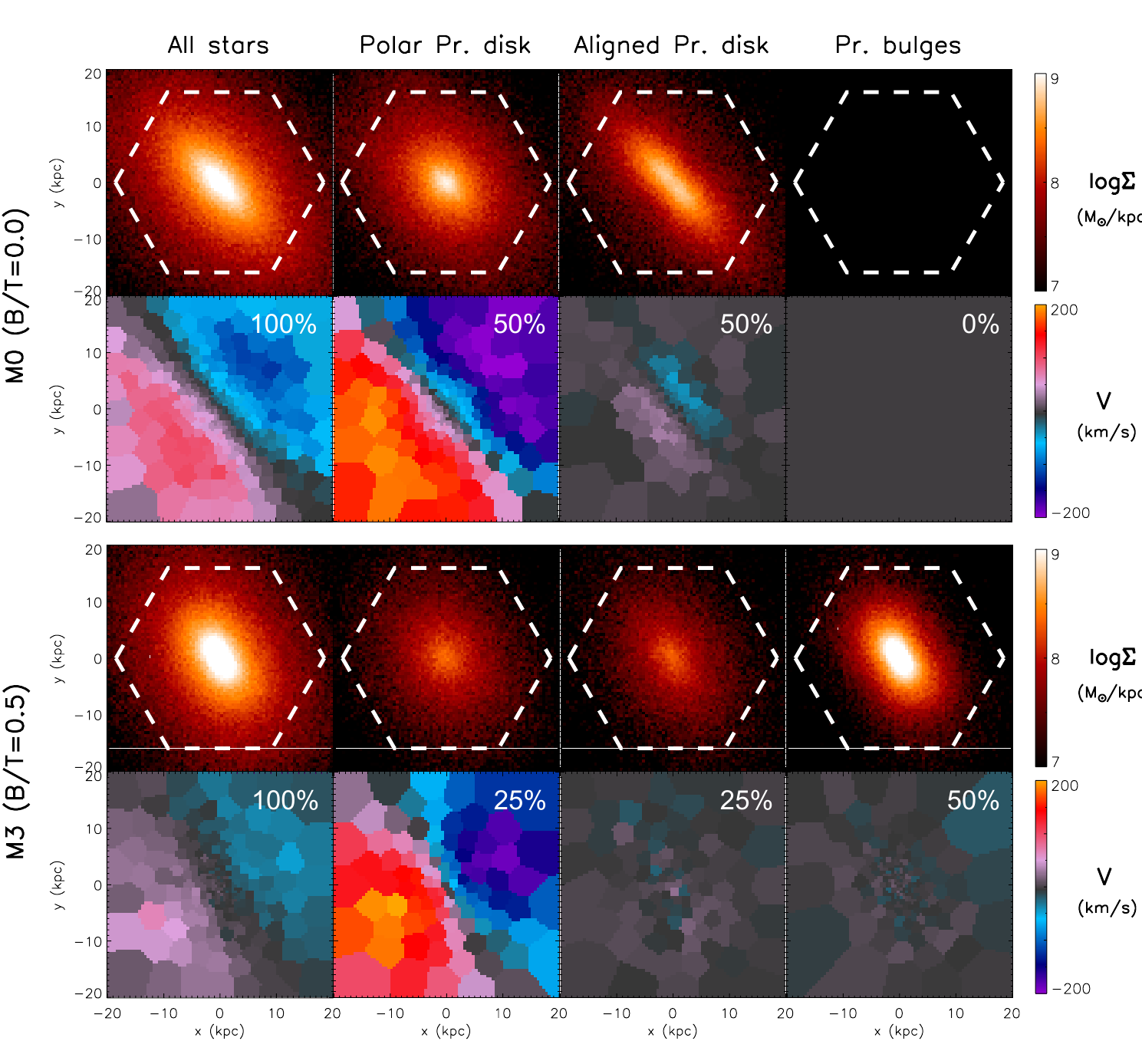}
\center
\caption{Same as Figure~\ref{fig:figure3}, for the remnants M0 (top two rows) and M3 (bottom two rows). From top to bottom: mass distribution $log\Sigma$ and mock line-of-sight velocity maps for different stellar populations in the remnant. From left to right: the first column shows all the stars of the merger remnant (``All stars''), the second column shows only the particles that initially belonged to the disk of the progenitor that was inclined by $90^{\circ}$ with respect to the orbital plane (``Polar Pr. disk'') and show strong prolate rotation in the remnant. The third column shows the same for the particles that initially formed the disk of the progenitor that was aligned with the orbital plane (``Aligned Pr. disk''), while the last column shows the rest of the stellar particles of the remnant, that initially formed the progenitors bulges (``Pr. bulges''). The mass fraction of each component with respect to the total stellar mass of the remnant is denoted at the top right corner.}
\label{fig:figure5}
\end{figure*}
\begin{figure*}
\center
\includegraphics[trim = 0.5cm 18cm 15.5cm 0cm,clip, width=18cm, ]{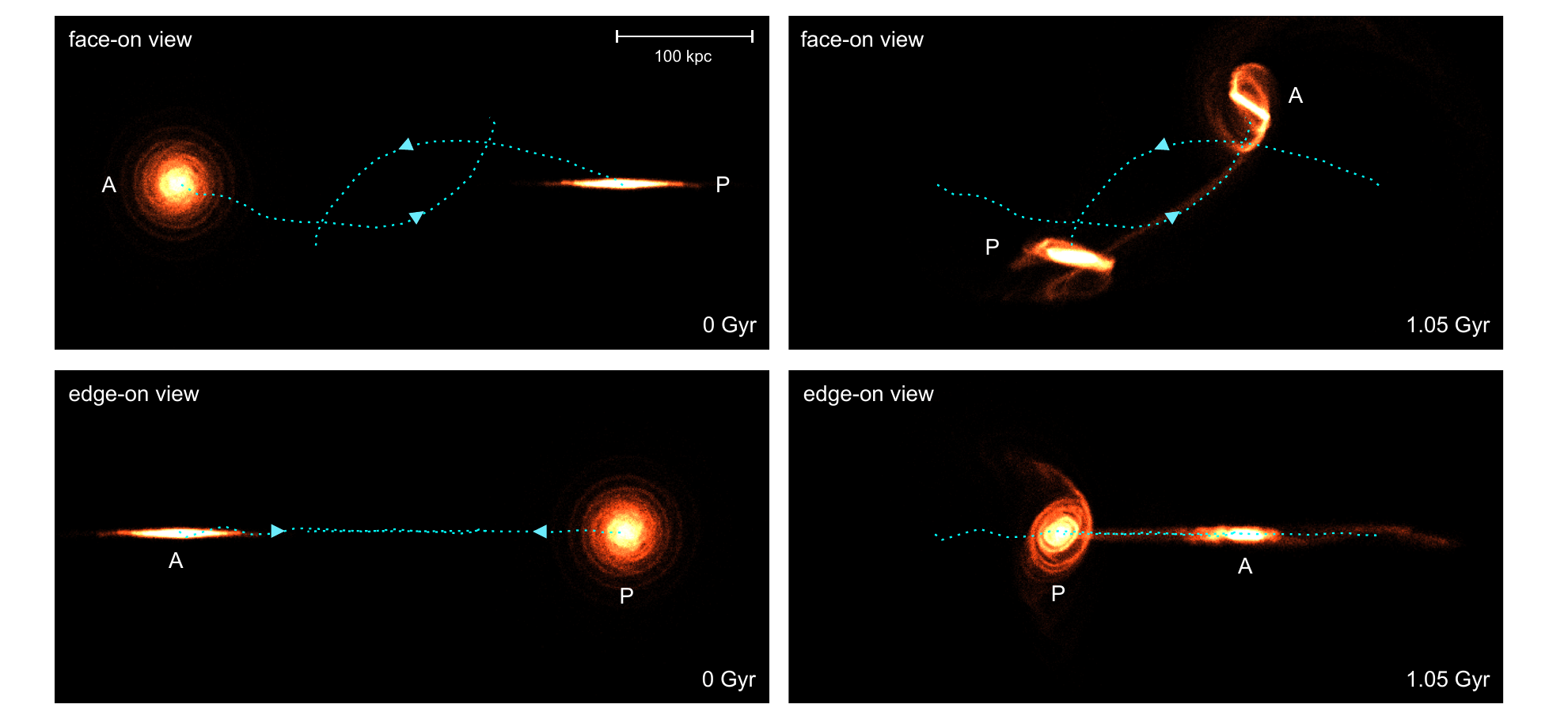}
\center
\caption{Face-on (top panels) and edge-on (bottom panels) projection of the orbital plane of the merger simulation M0. The aligned progenitor is denoted as ``A'', while the polar progenitor as ``P''. The panels on the left show in the initial set-up before the merger simulation starts (t=0 Gyr). The panels on the right show the two progenitors 1.05 Gyr later, at their first apocenter, where the aligned progenitor A has developed a strong, tidally induced bar (seen face-on at the top right panel), while this is not the case for the polar progenitor P (seen face-on at the bottom right panel). The merger trajectories of the two progenitors are overplotted. Images were created with glnemo2.\protect\footnotemark}
\label{fig:figure7}
\end{figure*}

The velocity dispersion shows two peaks along the minor photometric axis and a central dip for the bulgeless case. As the progenitor B/T ratio increases, the double peaks become weaker. Nonetheless, the CALIFA velocity dispersion maps shown in Figure~\ref{fig:figure1} have too low spatial resolution and relatively large uncertainties, that it is difficult to identify the existence of such substructures.

We additionally investigate the spatial structure of the higher order moments, ${h_3}$ and ${h_4}$, which are comparable to the skewness and the kurtosis of the LOSVD, respectively. \cite{Krajnovic_2008} found that ETGs show two different trends in their higher order kinematics: the first trend shows ETGs with a strong anticorrelation of ${h_3}$ and $V/\sigma$ with large values of $V/\sigma$, characteristic of a tail of low-velocity material in the LOSVD, and indicates the existence of a rotating disk component in the galaxy. The second trend shows an ${h_3}$ that is close to zero, which may show even positive correlation at intermediate $V/\sigma$, which is considered as the result of the existence of prolate rotation (minor axis contamination).

Figure~\ref{fig:figure4} shows that for all our simulated remnants ${h_3}$ and $V/\sigma$ show a positive correlation at low $V/\sigma$ values, especially for the remnants of bulge dominated progenitors (B/T\textgreater0.3). For the remnants of more disky progenitors (B/T\textless0.3) the distribution of ${h_3}$ and $V/\sigma$ shows a positive correlation at intermediate $V/\sigma$ and anticorrelating tails at higher $V/\sigma$ values ($V/\sigma$\textgreater 0.3).

A positive correlation between ${h_3}$ and $V/\sigma$ is indicative of a tail of high-velocity material in the LOSVD. In the case of triaxial ellipsoids, it has been shown to be caused by the superposition of box orbits and minor axis tube orbits in simulations of collisionless merger remnants \citep{Bendo_2000, Jesseit_2007}. In line with these findings, we confirm this feature in our triaxial and collisionless merger remnants and we additionally suggest that the amount of box versus minor axis tube orbits in the final remnant seem to be linked to the orbital structure of each progenitor, as the progenitor B/T ratio is influencing the resulting ${h_3}$ and $V/\sigma$ distribution of each remnant (Figure~\ref{fig:figure4}).

Similarly, the CALIFA data are challenging in order to measure robustly the higher order moments ${h_3}$ and ${h_4}$, however we report evidence for an ${h_3}$-$V/\sigma$ correlation for \object{LSBCF560-04} (Figure~\ref{fig:figure4}). By calculating the Pearson correlation coefficients $\rho$ of 10000 bootstrap realisations of ${h_3}$ and $V/\sigma$ with replacement, we find that the distribution of $\rho$ resembles a Gaussian distribution, with mean of $\bar{\rho}$=0.26 and $\sigma$=0.14, suggesting a weak positive correlation of ${h_3}$ and $V/\sigma$ for \object{LSBCF560-04}, which becomes stronger ($\bar{\rho}$=0.50, $\sigma$=0.20) at values $V/\sigma$ \textgreater 0.08. With higher sensitivity and better resolution instruments, such as MUSE \citep{Bacon_2010}, it will be possible and very interesting to to see if the prolate-rotating CALIFA galaxies reported here show similar signatures in their stellar velocity dispersion and their higher-order moments, as the ones resulting from the gas-poor merger formation history assumed here.

We also find that the triaxial merger remnants exhibit figure rotation on the orbital plane. The direction of the figure rotation is the same as the orbital direction of the merger. However, this tumbling motion is very slow, and for example in the case of simulation M3 the long axis of the remnant completes one full rotation in approximately 4.5 Gyr. This figure rotation is caused either by torques exerted from the triaxial dark matter halo, or/and by angular momentum transfer from the stellar component that is expelled during the merger and gets subsequently re-accreted onto the main stellar halo. Although it is beyond the scope of this paper to study in detail the effect of figure rotation, we note that if strong enough, it may alter the orbital structure of a galaxy and its observed kinematics \citep[e.g.][]{Heisler_1982, vanAlbada_1982, Wilkinson_1982, Statler_2004, Deibel_2011}.
In our case the figure rotation is slow and is not evident in our line-of-sight mock kinematic maps, as each remnant is projected with the orbital plane (which is also the plane of figure rotation) viewed face-on.\\

  \subsection{Origin of prolate rotation}
In order to understand better the origin of prolate rotation, we separate the LOSVD of the stellar particles in each remnant according to their formation origin, by selecting the particles that initially formed the disks and the bulges of the two progenitors.

Figure~\ref{fig:figure5} shows, for the remnant of two bulgeless progenitors M0, as well as for the remnant of two bulge dominated (B/T=0.5) progenitors M3, that the particles that initially formed the disk of the progenitor that was inclined by 90$^{\circ}$ with respect to the orbital plane of the merger (``Polar Pr. disk'') show strong prolate rotation in the final merger remnant, of an amplitude of $\sim$200 km/s for both simulation set-ups.
The particles that initially formed the disk of the progenitor whose disk was aligned with the orbital plane (``Aligned Pr. disk'') show very weak rotation of an amplitude of $\sim$10 km/s. The rest of the stellar particles that initially formed the bulges of the two progenitors (``Pr. bulges'') show no rotation in the final remnant (M3). 

This implies that depending on the B/T of the progenitors, the stellar population that accounts mainly for the prolate rotation may contribute significantly ($\sim$50\%) to the total mass of a remnant of disky progenitors (M0), while in the case of two bulge-dominated progenitors (M3) this component may be only $\sim$25\% of the total mass of the remnant.

The latter seems to be in agreement with findings for \object{NGC 4365}, a well-studied elliptical galaxy that shows prolate rotation in its outer parts with an amplitude of $\sim$60 km/s \citep{Davies_2001}. By constructing triaxial Schwarzschild dynamical models, \cite{vdB_2008} have shown that the stellar component of this galaxy that accounts for its prolate rotation, shows an amplitude of stellar rotation which is larger than $\sim$150 km/s, although contributing only $\sim$20\% to the total stellar mass of the \object{NGC 4365}.

In our simulations, stars that account for the prolate rotation retain the memory of their initial orientation prior to the merger, and for many Gyr of evolution after its progenitors have merged. These stars, that initially belonged to the ``polar'' progenitor's disk, show a similar amplitude of rotation in the remnants in all simulation set-ups, hence their mass fraction defines the amplitude of prolate rotation of the whole galaxy. 

Figure~\ref{fig:figure5} also shows that the aligned progenitor composes a very elongated, almost bar-like component in the final remnant, while the polar progenitor shows a less prolate shape. We find that this is true for all simulation set-ups, and for example for simulation M0, the triaxiality parameter for the aligned progenitor is T=0.87, while for the polar T=0.55 in the final merger remnant.

The different shapes of the two progenitors in the final merger remnant is a consequence of the different impact of their mutual tidal interactions on their morphology during their merger. Figure~\ref{fig:figure7} shows the two progenitors in simulation M0 in their initial set-up before the merger simulation starts (t=0 Gyr) and at their first apocenter (t=1.05 Gyr). While the two galaxies are almost identical initially, the aligned progenitor develops a strongly barred morphology soon after the first pericentric passage (top right panel of Figure~\ref{fig:figure7}). This is not the case for the polar progenitor, which retains its disky, although substantially disturbed, shape (bottom left panel of Figure~\ref{fig:figure7}).

Such tidally induced bars have been previously studied in detail to form after first pericentric passages of disk galaxies during encounters with a perturber- they are long-lived, and caused by angular momentum transfer from the stellar component of the galaxy to its halo \citep[e.g.][]{Gerin_1990, Aguerri_2009, Lokas_bar_2014, Lokas_bar_2016, MarValpuesta_2017}.
In our case, the tidally induced bar of the aligned progenitor is also long-lived and survives even after the coalescence of the two galaxies at t$\sim$2 Gyr, and until the final timestep of the simulation ($\sim$6 Gyr after coalescence) the aligned progenitor retains this bar-like, prolate shape in the final remnant.
 
We conclude that in the final merger remnant, the aligned progenitor retains memory of its shape before coalescence and is mainly responsible for the prolate shape of the remnant. On the other hand, the polar progenitor retains memory of its angular momentum before coalescence, and is responsible for the prolate rotation of the remnant.

We expect that our findings, combined with future, orbit-based dynamical models for the CALIFA sample of prolate rotators, can help towards better constraining their dynamical structure as well as their formation history, as the mass fraction of their prolate-rotating populations may provide important information on their assembly process and the nature of their progenitor galaxies.\\

\section{Summary and Discussion}
\label{sec:4}


We present evidence for 10 ETGs from the CALIFA Survey that show prolate rotation (rotation around the major photometric axis) in their stellar kinematics. This sample includes the discovery of 8 new prolate rotators, adding a significant fraction to the ($\sim$11) such cases of massive ETGs that have been reported so far in the literature. We additionally investigate their possible merger origin by studying the stellar kinematics of elliptical merger remnants using $N$-body simulations of major polar galaxy mergers. Our results can be summarised as follows:\\
\footnotetext{\url{http://projets.lam.fr/projects/glnemo2}}
\begin{itemize}

\item[(i)] Most of the ten prolate rotating galaxies presented here appear to belong to galaxy groups or clusters. Five of them are BCGs (Brightest Cluster Galaxies). Two of them show distinct minor axis dust lanes (\object{NGC 0810} and \object{NGC 5485}). Together with the main stellar body prolate rotation, minor axis dust lanes are an additional evidence for triaxiality \citep{Bertola_1978, Merritt_1983}. We suggest that the galaxies presented here are intrinsically triaxial systems.

\item[(ii)] Early-type galaxies with prolate rotation might be more common than previously thought. We have detected 6 prolate rotators out of 81 ETGs in the CALIFA kinematic Sub-sample of 300 galaxies as described in \cite{FalconBArroso_2016}. This corresponds to a volume-corrected fraction of $\sim$9\% of ETGs with prolate rotation. This unprecedentedly high fraction yields potential implications for ETG formation.

\item[(iii)] We investigate the merger scenario, according to which the central stellar body of a prolate rotator was formed by a major merger more than $\sim$10 Gyr ago. As for most of the CALIFA galaxies we see no evidence for oblate stellar rotation (rotation around the short axis), we suggest that their merger formation must have been gas-poor. We thus perform a set of $N$-body simulations of major polar mergers of disk galaxies and investigate the dynamical structure of their resulting remnants. We find that such remnants exhibit highly prolate shapes with strong prolate rotation that depends on the B/T (bulge-to-total stellar mass) ratio of their progenitor galaxies. The higher this ratio is, the lower the amplitude of prolate rotation in the resulting ETGs.

\item[(v)] By constructing mock IFU observations of their stellar kinematics, we find that all the simulated merger remnants show a double peak on their line-of-sight (LOS) velocity dispersion profile along the minor axis. As the B/T of the progenitors increases, the double peaks become weaker. We also find a positive correlation between their LOS velocity and the higher-order-moment $h_3$. This is in contrast with what is observed for most ETGs with oblate rotation \citep[e.g.][]{Krajnovic_2008}. Better quality observations are now needed in order to confirm the presence of such kinematic features in the observed prolate rotators.

\item[(vi)] We show that the prolate rotation in each simulated merger remnant originates from the progenitor galaxy that had its disk in an orthogonal orientation with respect to the orbital plane prior to the merger, while the prolate shape of the remnant originates mainly from the progenitor that had its disk aligned with the orbital plane. The latter is a consequence of the aligned progenitor developing a tidally induced bar before coalescence.\\

\end{itemize}


\subsection{Discussion}
Here we have investigated the polar merger origin of prolate rotating ETGs, using mergers of two identical disk galaxies with orthogonal disk orientations prior to the merger. We note that such orientations may be infrequent, and one would expect that varying the initial relative inclinations of the two disks would change the resulting kinematics of the remnant. 

The origin of prolate rotation has also been studied in simulations of dwarf galaxy mergers in orthogonal disk orientations \citep{Lokas_2014}, which involved, however, progenitors on radial orbits, with their disks inclined by 45$^{\circ}$ with respect to their orbital plane. This formation scenario was suggested in order to explain the prolate rotation that was recently observed in the dwarf spheroidal galaxy Andromeda II \citep{Ho_2012}, the first dwarf galaxy\footnote{See also \cite{Kacharov_2016}, for their recent discovery of prolate rotation in another dwarf galaxy.} observed to show such a feature in the form of a stellar stream rotating around its major projected axis, and which was interpreted as evidence of a past major merger between two dwarf galaxies \citep{Amorisco_2014}. 

By studying this merger scenario, \cite{Ebrova_2015} found that mergers between disky dwarfs are needed to explain the prolate rotation in the resulting dwarf galaxy remnants, which can be accounted for by a variety of inclinations of the progenitor disks and orbital plane orientations. 

In line with these findings for dwarf galaxies, here we have studied the scenario of polar mergers for the case of massive ETGs with prolate rotation, for which we would expect that they may as well result from a variety of initial orientations of their progenitors disks, and that our fine-tuning of a strictly orthogonal orientation should not be considered as a limitation to their predicted rate of occurrence. 

In this work we have investigated only the effect of the B/T fraction of each progenitor, however we note that various other parameters may influence the resulting kinematics of the merger remnant. Such parameters include not only the initial orbital parameters of the merger, but also other structural parameters of the progenitor disks, as well as their relative mass fraction. Such a further investigation remains beyond the scope of the present work, however seems now crucial towards a better understanding of the formation origin of prolate rotators.

In our simulations we find prolate remnant ETGs with a positive correlation between their LOS velocity and the higher-order-moment $h_3$. The quality of the CALIFA data is not sufficient to measure robustly the higher-order moments in order to compare with findings from our merger simulations. These CALIFA ETGs would require further study with higher resolution observations from next generation IFU instruments (e.g. MUSE, \citealt{Bacon_2010}) in order to assess further the implications of our findings.

This work, combined with future, higher quality observations of the sample of galaxies presented and orbit-based dynamical modeling, will give better insights into the dynamical nature of this special type of rotators. Finally, our findings, combined with results from cosmological simulations, will help towards setting better constraints on the formation origin as well as the rate of occurrence of prolate rotation in massive ETGs.

\begin{acknowledgements}
We are grateful to the anonymous referee for the constructive suggestions for improvement of our work, as well as to Hector Hiss, Ivana Ebrov\'a, Davor Krajnovi\'c, Eric Emsellem and Sebasti\'an S\'anchez for useful discussions and contributions. This study makes use of the data provided by the Calar Alto Legacy Integral Field Area (CALIFA) survey \url{http://califa.caha.es}. Based on observations collected at the Centro Astron\'omico Hispano Alem\'an (CAHA) at Calar Alto, operated jointly by the Max-Planck-Institut f\"ur Astronomie and the Instituto de Astrof\'isica de Andaluc\'ia (CSIC). JALA was funded from the grant AYA2013-43188-P, and JFB from grant AYA2016-77237-C3-1-P, by the Ministerio de Economia y Competitividad (MINECO). AT and GvdV acknowledge financial support from the DAGAL network from the People Programme (Marie Curie Actions) of the European Union's Seventh Framework Programme FP7/2007- 2013/ under REA grant agreement number PITN-GA-2011-289313. The numerical simulations used in this work were performed on the THEO cluster of the Max-Planck-Institut f\"ur Astronomie at the Rechenzentrum in Garching. Glnemo2 copyright: Jean-Charles Lambert. Glnemo2 was developed at CeSAM/LAM.

\end{acknowledgements}

\bibliographystyle{aa}


\end{document}